\documentclass[journal]{rmaa}

\title{The exo-planetary system of 55 Cancri\\ and the Titius-Bode Law}

\author{
Arcadio Poveda\altaffilmark{1} 
and Patricia Lara\altaffilmark{2}}

\altaffiltext{1}{Instituto de Astronom\'ia, Universidad Nacional Aut\'o\-noma de M\'exico, Mexico.}

\altaffiltext{2}{Facultad de Ciencias, Universidad Nacional Aut\'onoma de M\'exico, Mexico.}

\shortauthor{Poveda \& Lara} 

\shorttitle{55 Cancri and the Titius-Bode Law}

\listofauthors{A.~Poveda \& P.~Lara}

\indexauthor{Poveda, A.}
\indexauthor{Lara, P.}

\fulladdresses{
\item Arcadio Poveda: Instituto de Astronom\'ia, Universidad Nacional Aut\'onoma de M\'exico, 
Apdo. Postal 70-264, 04510 M\'exico, D.~F. Mexico (poveda@servidor.unam.mx).
\item Patricia Lara: Facultad de Ciencias, Universidad Nacional Aut\'onoma de M\'exico, Av. 
Universidad 3000, Circuito Exterior S/N, Ciudad Universitaria, 04510 M\'exico, D.~F. Mexico 
(path\_6@yahoo.com).}

\ReceivedDate{2008 February 13}
\AcceptedDate{2008 February 26}
\SetYear{2008}

\resumen{El reciente descubrimiento de un quinto planeta ligado a 55 Cancri 
(Fischer et al. 2007) nos ha motivado a investigar si este sistema exo-planetario 
se ajusta a alguna una forma de la ley de Titius-Bode (TB). Encontramos que 
una simple relaci\'on TB exponencial reproduce muy bien los cinco semiejes 
mayores observados siempre y cuando se asigne el n\'umero 6 al planeta con 
el semieje m\'as grande. Esta forma de contar deja un vac\'io en la posici\'on 
$n=6$, una situaci\'on curiosamente reminiscente a la ley TB en nuestro 
propio sistema planetario, antes del descubrimiento de Ceres. La aplicaci\'on 
de una ley T-B exponencial a 55 Cancri nos permite predecir la existencia 
de un planeta con $a \approx 2.0$~AU y con un per\'iodo de $P \approx 1130$~d\'ias 
localizado en la gran brecha entre $a=0.781$~AU ($P=260$~d\'ias) y $a=5.77$~AU 
($P=5218$~d\'ias) correspondientes a los dos m\'as grandes per\'iodos observados. 
Con menos certeza, tambi\'en predecimos un s\'eptimo planeta en $a \approx 15$~AU 
con $P \approx 62$~a\~nos.}

\abstract{The recent discovery of a fifth planet bound to 55 Cancri (Fischer 
et. al 2007) motivated us to investigate if this exo-planetary system fits 
some form of the Titius-Bode (TB) law. We found that a simple exponential TB 
relation reproduces very well the five observed major semi-axis, provided we 
assign the orbital $n=6$ to the largest $a$. This way of counting leaves 
empty the position $n=5$, a situation curiously reminiscent of TB law in 
our planetary system, before the discovery of Ceres. The application of an 
exponential TB relation to 55 Cancri allows us to predict the existence of a 
planet at $a \approx 2.0$ AU with a period of $P \approx 1130$ days located 
within the large gap between $a=0.781$~AU ($P=260$ days) and $a=5.77$~AU 
($P=5218$ days). With less certainty, we also predict a seventh planet at 
$a \approx 15$~AU, with $P \approx 62$ years.\smallskip}

\addkeyword{planetary systems}
\addkeyword{planets and satellites: general}
\addkeyword{stars: individual (55 Cancri)}

\begin{document}

\maketitle

\section{Introduction}

\vspace{0.1cm}

The two hundred year old saga of the Titius-Bode law is well known
(see Nieto 1972 for a well documented review). Since the discovery
by Bode, in 1782, that the Titius relation ``predicted''  the major
semi-axis of Uranus, a frantic search for ``the lost planet'' at
position $n=5$ was initiated at various European observatories.
The discovery of Ceres by Piazzi on the night of January 1st, 1801, 
with the major semi-axis predicted by TB, and the fact
that the satellites of Jupiter, Saturn and Uranus also follow a TB
relation, initiated a debate about the meaning of TB which is still
alive nowadays. Is the TB law a matter of chance (Lynch 2003; Dubrulle \&
Graner 1994; Neslu\v san 2004)? Is it consequence of the early
physical conditions in the protoplanetary disk (Graner \& Dubrulle
1994)? Is it a reflection of a process of dynamical relaxation in a
system of planets subject to their mutual gravitational
perturbation (Hayes \& Tremaine 1998; Hills 1970; Ovenden 1975)?

Because of the previous considerations and in view of the growing
number of multiple exo-planetary systems, we decided to check
whether the 55 Cancri system, for which a fifth planet has been
recently announced (Fischer et al. 2007), follows the TB law.


\begin{table}[!t]
\vspace{0.15cm}
\centering
\caption{Properties of 55 Cancri}
\small
\begin{tabular}{lcc}
\toprule
Apparent visual magnitude   & & $5.96$ \\
Hipparcos parallax          & & $79.8 \pm 0.84$ mas \\
Distance                    & & $12.5 \pm 0.13$ parsecs \\
Absolute visual magnitude   & & $5.47$ \\
Effective Temperature       & & $5234\pm30$K \\
Rotation velocity $v\sin i$ & & $2.4\pm0.5$ km s$^{-1}$ \\
Luminosity                  & & $0.6 \,\, L_\odot$ \\
Spectral type               & & G8V/K0V \\
Mass                        & & $0.94\pm0.05 \,\, M_\odot$ \\
\bottomrule
\end{tabular}
\vspace{0.4cm}
\end{table}


The traditional TB relation is essentially a geometric progression
in the number $n$, the running number of a planet according to its
distance to the central star. This geometric relation can be
represented by an exponential in $n$.

We tried to represent our planetary system by an exponential in $n$
and found a good fit. Having verified that an exponential fit for
the Solar System was a good approximation we tried to represent the 
55 Cancri system also by an exponential TB. The exponential fit to 
the 55 Cancri system was very good (with a coefficient of correlation 
$R^2=0.997$) when we assigned the number $n=6$ to the largest 
major semi-axis observed. The vacancy left at $n=5$ leads us to 
propose the existence of a new planet with a major semi-axis $a \approx 2$~AU.


\begin{table*}[!t]
\vspace*{0.15cm}
\centering
\setlength{\tabnotewidth}{\columnwidth}
\tablecols{6}
\caption{Observed Properties of the 55 Cancri Exo-planetary System\tabnotemark{*}}
\small
\begin{tabular}{lccccc}
\toprule
\multicolumn{1}{c}{$n$} & 1 & 2 & 3 & 4 & 5 \\
\midrule
Year of discovery & 2004 & 1996 & 2002 (1) & 2007 & 2002 (2) \\
\midrule 
Observed semiaxis major (AU) & $0.038$ & $0.115$ & $0.24$ & $0.781$ & $5.77$  \\
& $\pm 1.0 \times 10^{-6}$ & $\pm 1.1 \times 10^{-6}$ & $\pm 4.5 \times 10^{-5}$ & $\pm 0.007$ 
& $\pm 0.11$ \\
\midrule 
Period (days) & $2.817$ & $14.651$ & $44.344$ & $260$ & $5218$ \\
& $\pm 1 \times 10^{-4}$ & $\pm 7 \times 10^{-4}$ & $\pm 7 \times 10^{-3}$ & $\pm 1.1$ & $\pm 230$ \\
\midrule 
$M \sin i$ (Jovian masses) & $0.034$ & $0.824$ & $0.169$ & $0.144$ & $3.835$ \\
& $\pm 0.0036$ & $\pm 0.007$ & $\pm 0.008$ & $\pm 0.04$ & $\pm 0.08$ \\
\bottomrule
\tabnotetext{*}{Taken from Fischer et al. 2007.}
\end{tabular}
\vspace*{0.3cm}
\end{table*}


\section{ The 55 Cancri System}

The star 55 Cancri (55 Cnc = HD 75732 = HR3522 = HIP 43587) is a
well observed nearby star; in Table~1 we list some of its parameters
taken from the paper by Fischer et al. (2007). In Table~2 the observed 
parameters for its planetary system are listed, including the year 
of discovery of each planet. Note in this table the enormous spacing 
between planets $n=4$ and $n=5$, whose major semi-axes are, respectively, 
less than 1 AU and more than 5 AU, and whose periods are 260 days 
and 5218 days.


\begin{figure}[!t]
\vspace{0.05cm}
\centering
\includegraphics[width=0.99\columnwidth]{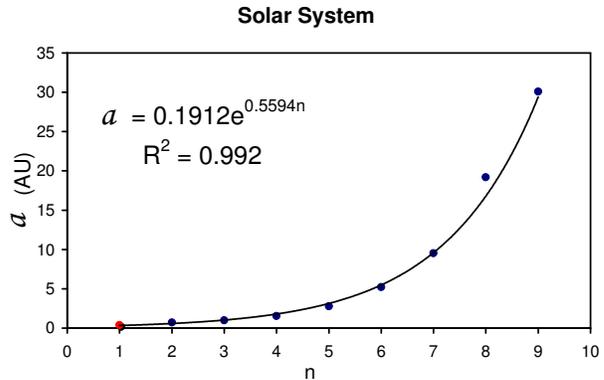}
\vspace{-0.3cm}
\caption{An exponential TB fit to the Solar System. This fit includes 
major semi-axis from Venus ($n=2$) up to Neptune ($n=9$). Note that the 
extrapolation of equation to $n=1$, gives $a(1)=0.335$~AU, close to the 
observed value $a=0.387$.}
\vspace{0.4cm}
\end{figure}


\section{ The Titius-Bode Law}

The equation
\begin{equation}
a = 0.4 + 0.3 \times 2^n \qquad (a \ {\rm in \ AU})
\end{equation}
\noindent represents the classical Titius-Bode law. Note the peculiar 
ordering system: Mercury corresponds to $n=-\infty$, Venus to $n=0\dots$

In the equation
\begin{equation}
a = 0.1912 \quad e^{0.5594n} \, .
\end{equation}
\noindent we present our best exponential fit to the Solar System
excluding Mercury and Pluto, but including Uranus and Neptune. Our
exponential TB fit to the Solar System excludes Mercury because in
the traditional TB relation not only it is given an orbital number
$n=-\infty$, devoid of any physical meaning, but also the value of 
the constant (0.4) is arbitrarily chosen to give the approximately 
correct values of $a$ for Mercury and the Earth. At the other end 
of our planetary system we exclude Pluto not only because of its 
pathological orbit, but also because we do not know if it is an object 
from the Kuiper belt, captured into the region of the outer planets, 
or a satellite ejected from Neptune; in any case its present orbit 
has followed a dynamical evolution different from that of the rest 
of the planets.


\begin{figure}[!t]
\vspace{0.08cm}
\centering
\includegraphics[width=0.99\columnwidth]{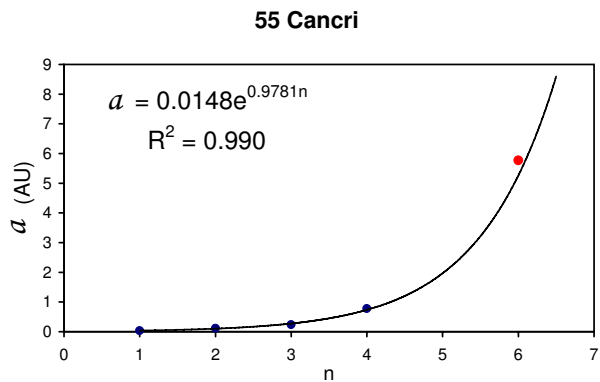}
\caption{An exponential TB fit to the four closest planets in the 55 Cancri 
system. Note that the extrapolation to $n=6$ corresponds to a major semi-axis 
close to the observed one.}
\vspace{0.3cm}
\end{figure}



\begin{figure}[!t]
\centering
\includegraphics[width=0.99\columnwidth]{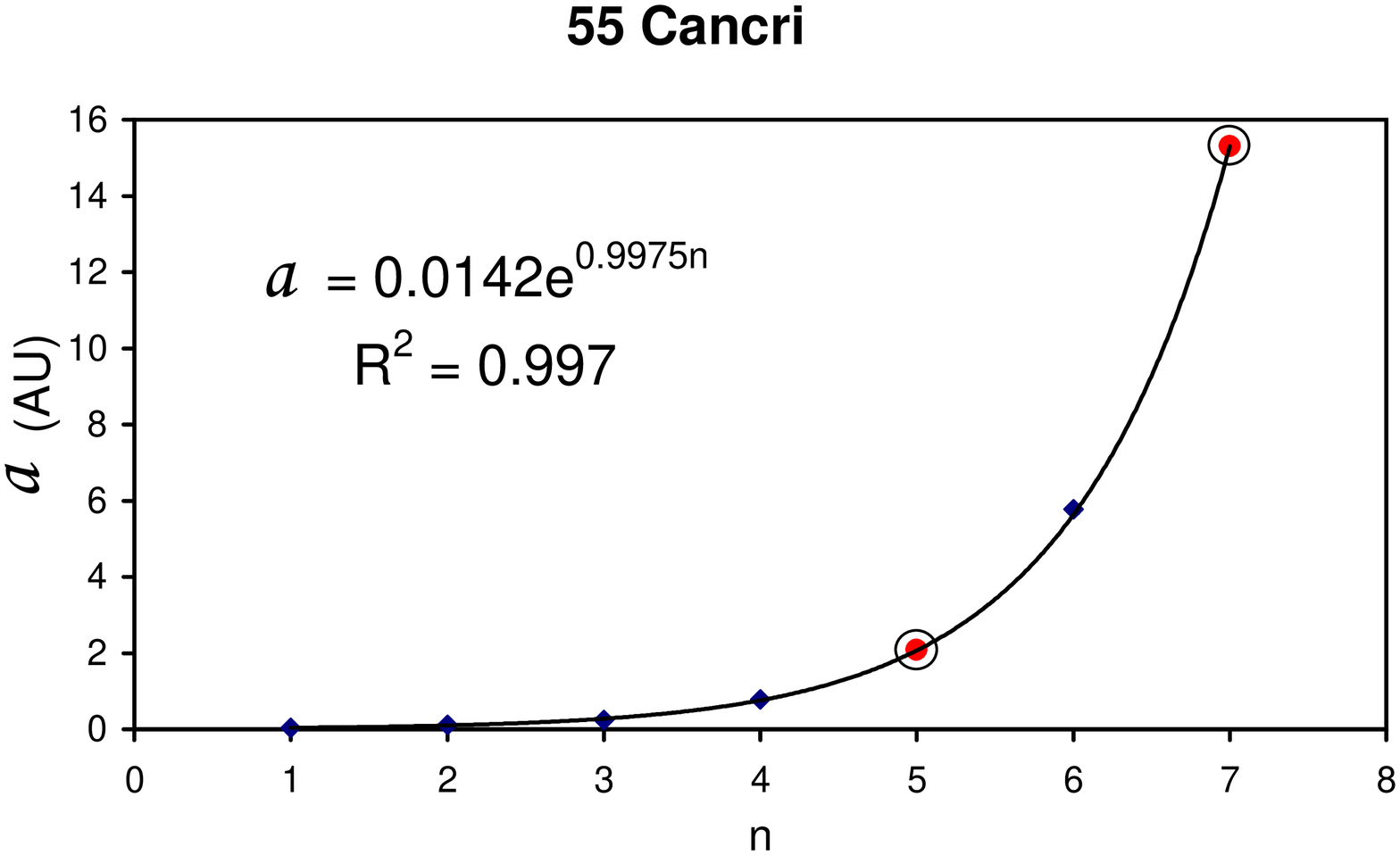}
\caption{The TB exponential fit to the 5 observed planets of 55 Cancri, where 
we count the farthest observed one as $n=6$. Planets $n=5$ and $n=7$, predicted 
by TB, are shown as open circles.}
\vspace{0.3cm}
\end{figure}


In Figure~1 we plot equation (2), and we mark the position of the 
orbital $n=1$. Note that although equation (2) does not include the 
value of $a$ for Mercury, it predicts orbitals close to the observed 
ones up to Neptune, $n=9$. The traditional TB relation gave a good 
fit up to Uranus, but a very poor one for Neptune. Our exponential 
TB relation gives a poor representation for Uranus, but a good one 
for Neptune.

\vspace{0.15cm}

\section{The Titius-Bode and the 55 Cancri exo-planetary system}

If we plot, as we do in Figure~2, the major semi-axis of the five 
known planets versus $n$, we note that the more distant planet, $n=5$ 
falls very far from any reasonable exponential fit to the 4 closest 
planets ($n=1, 2, 3, 4$). However, if we assume that planet 5 has 
not been discovered, and we fit the first 4 planets with an exponential 
we see that the extrapolation of this relation to $n=6$ ($a \approx 5.24$~AU) 
turns out to be close to the observed value ($a = 5.77$~AU). 
Figure~3 shows the best fit for the distribution of the observed major 
semi-axes of the first four planets and the fifth one, now occupying 
orbital number 6.

This fit is
\begin{equation}
a = 0.0142 \quad e^{0.9975n} \quad  (R^2 = 0.997)
\end{equation}


\begin{table*}[!t]
\vspace*{0.15cm}
\centering
\caption{The Titius-Bode fit to the 55 Cancri System and the Two New Planets}
\setlength{\tabnotewidth}{\columnwidth}
\tablecols{8}
\small
\begin{tabular}{lcccccccc}
\toprule
\multicolumn{1}{c}{$n$} & 1 & 2 & 3 & 4 & 5 & 6 & 7\\
\midrule 
Observed semiaxis major (AU) & $0.038$ & $0.115$ & $0.24$ & $0.781$ &  & $5.77$  & \\
& $\pm 1\times 10^{-6}$ & $\pm 1.1 \times 10^{-6}$ & $\pm 4.5 \times 10^{-5}$ & $\pm 0.007$ 
& & $\pm 0.11$\\
\midrule
Titius-Bode (AU) & 0.039 & 0.104 & 0.283 & 0.768 & \bf{2.08} & 5.643 & \bf{15.3} \\
\midrule 
Period (days) & 2.817 & 14.651 & 44.344 & 260 & \bf{1130} & 5218  & \bf{22530} \\
& $\pm 1 \times 10^{-4}$ & $\pm 7 \times 10^{-4}$ & $\pm 7 \times 10^{-3}$ & $\pm 1.1$ & & $\pm 230$ & \\
\bottomrule
\vspace*{0.3cm}
\end{tabular}
\end{table*}


This equation ``predicts'' the existence of a fifth planet at $a \approx 2$~AU 
and, with less certainty, a seventh one at $a \approx 15$~AU.

In Table~3 we list the observed elements of 55 Cancri system as well 
as the TB fit and the two new planets predicted by the Titius-Bode law.

\vspace{0.2cm}

\section{Conclusions}

In the present paper we showed that the exponential Titius-Bode law 
holds for the 4 closest planets of 55 Cancri and that its extrapolation 
fits well the major semi-axis of the fifth planet, provided it is 
assumed that it occupies the sixth orbital.

\vspace{0.1cm}

The Titius-Bode law is valid for the exo-planetary system 55 Cancri, and 
may be valid for other exo-planetary systems as well.

\vspace{0.1cm}

Having found another planetary system where the Titius-Bode law is valid 
makes it rather unlikely that it is due to chance. 

\vspace{0.1cm}

The Titius-Bode law allows us to predict two new planets for the 55 Cancri 
system:
\begin{eqnarray}
a \approx ~2.0~ {\rm AU } & \qquad P \approx ~3.1~ {\rm years}\nonumber \\
a \approx 15.0~ {\rm AU } & \qquad P \approx 62~ {\rm years}\nonumber
\end{eqnarray}

The existence of two hot Jupiter-like planets ($n=1, 2$), in this system 
opens the problem of how to understand the persistence of the Titius-Bode 
law against the phenomenon of planet migration.

\vspace{0.27cm}

The validity of TB for the 55 Cancri exo-planetary system does not yet 
help to understand the physics behind it. However, it may help to discover 
new planets by paying special attention to periodic signals in the radial 
velocities at values close to the predicted periods.

\adjustfinalcols


\begin{thebibliography}{}

\bibitem{} Dubrulle, B., \& Graner, F. 1994, A\&A, 282, 269

\bibitem{} Fischer, D.~A., et al. 2007, ApJ, submitted (astro-ph/07123917)

\bibitem{} Graner, F., \& Dubrulle, B. 1994, A\&A, 282, 262

\bibitem{} Hayes, W., \& Tremaine, S. 1998, Icarus, 135, 549

\bibitem{} Hills, J.~G. 1970, Nature, 225, 840

\bibitem{} Neslu\v san, L. 2004, MNRAS, 351, 133

\bibitem{} Nieto M.~M. 1972, The Titius-Bode Law of Planetary Distances: Its History and 
Theory (Oxford: Pergamon Press)

\bibitem{} Ovenden, M.~W. 1975, Vistas in Astron., 18, 473

\end{thebibliography}
\end{document}